\documentclass[11pt]{article}    
\usepackage{graphicx}

\usepackage{geometry}
\usepackage{mathptmx}      
\usepackage{latexsym,amsmath,amssymb}
\usepackage{subfigure,multirow}
\usepackage{caption}
\usepackage{makecell,booktabs,footnote}
\usepackage[table]{xcolor}
\usepackage{bm}
\usepackage{hyperref}
\usepackage[ruled,vlined]{algorithm2e}
\usepackage{cite}
\usepackage{threeparttable}
\usepackage{url}
\usepackage{natbib}
\setcitestyle{authoryear,round}
\newcommand{\tabincell}[2]{\begin{tabular}{@{}#1@{}}#2\end{tabular}}

\begin{document}

\title{Bayesian Variable Selection via Hierarchical Gaussian Process Model in Computer Experiments}
\author{  Xiao Yao$^1$ \quad Ning Jianhui$^2$  \quad Qin Hong$^{1,2}$  \\
{\small$^1$ School of Statistics and Mathematics, Zhongnan University of Economics and Law, Wuhan, China}\\
{\small$^2$ School of Mathematics and Statistics, Central China Normal University, Wuhan, China}}

\date{}

\maketitle

\begin{abstract}
Identifying the active factors that have significant impacts on the output of the complex system is an important but challenging variable selection problem in computer experiments. In this paper, a Bayesian hierarchical Gaussian process model is developed and some latent indicator variables are embedded into this setting for the sake of labelling the important variables. The parameter estimation and variable selection can be processed simultaneously in a full Bayesian framework through an efficient Markov Chain Monte Carlo (MCMC) method---Metropolis-within-Gibbs sampler.  The much better performances of the proposed method compared with the related competitors are evaluated by the analysis of simulated examples and a practical application.

\noindent \textbf{keywords:} Computer experiments; Variable selection; Bayesian Gaussian process model; Normal mixture prior; Markov Chain Monte Carlo 
\end{abstract}

\section{Introduction}
As computer techniques rule our lives, computer  simulator  experiments have never been more important. They simulate complex physical phenomena/systems that are difficult (expensive, time-consuming) or even impossible to study via coded mathematical models \citep{santner2018}. Computer experiments frequently have complicated nonlinear input-output relationship with long simulation time, hence surrogate models / emulators / metamodels are employed to replace the computer simulators to analyse the resulting data \citep{fang2006}.  In addition,  computer experiments often involve large number of input variables, for example, engineering design choices, environmental conditions, uncertain model parameters and so on.  Thus it is important to identify those variables (relatively few)  having significant impacts on the response, which are termed as ``active" variables. This process is also known as \textit{variable selection} in statistics.  

A common  surrogate model in computer experiments is Gaussian process (GP) model, also named as ``Kriging" model,  which was firstly proposed by South African geologist D. G. Krige and then was systematically introduced to computer experiments by \citet{sacks1989a}. There are two categories of strategies about variable selection for Gaussian process model in computer experiments. The first class is focused on selecting the active variables (terms) in the mean function of Gaussian process model. Many novel methods, such as blind Kriging \citep{joseph2008}, penalised blind Kriging \citep{hung2011} and stochastic searched blind Kriging \citep{huang2020} were proposed to deal with this issue. However, these researches particularly work when strong long-term trend exists.   

Another family of identifying important variables for Gaussian process model is mainly concentrated on the part of modelling local deviations, i.e., the correlation structure of stationary Gaussian process. This idea stems from \cite{Welch1992}. They used the likelihood to select the important variables via the stepwise algorithm. However the likelihood of Gaussian process model is often unstable and usually leads to fairly large variance of the maximising likelihood estimator (MLE).    \cite{li2005} considered a penalised likelihood function to estimate the covariance parameters and identify the active variables. Furthermore,   the framework of Bayesian maximum a posterior (MAP) with Jeffery's non-informative prior for the covariance parameters  in Gaussian process model was employed  by \cite{deng2012}. This method could provide more accurate prediction with lower computational cost compared with penalised method. Another procedure for performing variable selection based on Gaussian process model via the framework of Bayes was proposed by \cite{linkletter2006}.  This method has two key features: one is the ``slab and spike" prior for the correlation parameters  and the other is  an augmented inert input that creates a reference distribution to identify the active variables.  


In this article, we develop a new variable selection procedure for Gaussian process model in computer experiments. We propose to construct  a Bayesian hierarchical GP model with a normal mixture prior for the correlation parameters and to use a latent indicator variable to ``label" whether the corresponding input being active or inactive.  This research offers several major contributions to the literature. First,  we introduce the normal mixture prior for the correlation parameters to the Bayesian hierarchical GP model and  conduct a simultaneous parameter estimation and variable selection in a full Bayesian context. What's more, an efficient Markov Chain Monte Carlo (MCMC) method is employed to deal with the posterior inference. Numerical results including simulated examples and real applicaation indicate that the proposed method performs better than the related competitors. 

The rest of this paper is organised as follows. Section 2 describes the Gaussian process model in computer experiments. In section 3, Bayesian hierarchical GP model with normal mixture prior is introduced and the posterior inference is done. Section 4 gives the numerical results. Section 6 concludes this paper. 
 
\section{Gaussian process model}

Suppose that $\mathcal{P}=\{\bm x_1,\cdots, \bm x_n\}$ are design points in the $d-$dimensional experimental domain $D$, and that $\bm y=(y_1, \cdots, y_n)^{\mathrm{T}}$ are responses. The model can be formed as 
\begin{equation}\label{eq:model}
	y(\bm x)= \mu + Z(\bm x)
\end{equation}
where $Z(\bm x) $ is a Gaussian Process with mean $E[Z(\bm x)]=0$ and $n \times n$ covariance matrix $\sigma^2R$. The $(i,j)$th element in correlation matrix $R$ is the correlation $r(\bm x_i, \bm x_j)$ between two experimental points. Model \eqref{eq:model} is often called as ordinary Kriging.

The correlation $r(\bm x_i, \bm x_j)$ is a positive semidefinite function with $r(\bm x_i, \bm x_i)=1$ and $r(\bm x_i, \bm x_j)=r(\bm x_j, \bm x_i)$. Usually, the Gaussian correlation function
\begin{equation}\label{eq:gscorr}
	r(\bm x_i, \bm x_j)=\exp\left\{-\sum_{k=1}^d\theta_k|x_{ik}-x_{jk}|^2\right\},
\end{equation} 
is adopted  by researchers and practitioners \citep{ba2012}.  The value of each parameter $\theta_k, k=1,\cdots,d$ reflect the importance of the corresponding $k$-th variable.  When $\theta_k$ equals to $0$,  the $k$-th variable is inactive; otherwise, it is active when $\theta_k$ is far away from $0$.

The parameters $\mu, \sigma^2, \bm\theta=(\theta_1,\cdots,\theta_d)^\mathrm{T}$ can be estimated by maximising the log-likelihood function (up to a constant):
\begin{equation}\label{eq:likelihood}
	l(\mu.\sigma^2, \bm \theta)\propto -\frac{1}{2}\left\{n\log(\sigma^2) +\log(\det R(\bm \theta)) +\frac{1}{\sigma^2}(\bm y-\bm 1_n \mu)^\mathrm{T}R^{-1}(\bm \theta)(\bm y-\bm 1_n \mu) \right\},
\end{equation}  
where $\bm 1_n$ is a vector whose elements are all $1$. Given the parameters $\bm \theta$, the MLE of $\mu$ is $\hat{\mu}=(\bm 1_nR(\bm \theta)\bm1_n)^{-1}\bm 1_nR(\bm \theta)\bm y$, and the MLE of $\sigma^2$ is $\hat{\sigma}^2=\frac{1}{n}(\bm y-\bm 1_n \hat{\mu})^\mathrm{T}R^{-1}(\bm \theta)(\bm y-\bm 1_n \hat{\mu})$.  Substituting $\hat{\mu}$ and $\hat{\sigma}^2$ into the log-likelihood function \eqref{eq:likelihood}, the  MLE  of $\bm \theta$ is calculated by numerically maximising  $-\frac{1}{2}[n\log(\hat{\sigma}^2)+\log(\det R(\bm\theta))]$, denoted as $\hat{\bm \theta}$. The details can be referred to \cite{fang2006, santner2018}. 

The prediction at an untried/test experimental point $\bm x$ can be obtained 
\begin{equation}\label{eq:km}
	\hat{y}(\bm x)=\hat{\mu}+r^\mathrm{T}(\bm x)R^{-1}(\hat{\bm \theta})(\bm y-\bm 1_n \hat{\mu}),
\end{equation}
where $r^\mathrm{T}(\bm x)=(r(\bm x, \bm x_1), \cdots, r(\bm x, \bm x_n))$. In the meanwhile, the mean square error of prediction at the test point $\bm x$ can be given as 
\begin{equation}\label{eq:ks}
	s^2(\bm x)= \hat \sigma^2\left[1-r^TR^{-1}r+\frac{(1-\bm 1_n^TR^{-1}r)^2}{\bm 1_n^TR^{-1}\bm 1_n} \right].
\end{equation} 

\section{Bayesian framework for variable selection}
\subsection{Stochastic search Gaussian process model}
In the previous sections, it is known that the correlation parameters $\theta_k$'s reflect the importance of the corresponding variables, i.e., selecting a set of active variables is equivalent to setting those $\theta_k$'s corresponding to non-selected variables to 0. In order to satisfy this purpose, the normal mixture hierarchical framework for variable selection in \citet{george1993} is embedded  into GP model.  
This hierarchical setting is famous as Stochastic Search Variable Selection (SSVS) scheme and is also used for uncertainty quantification studies \citep{chen2017} and for RBF global optimisation \citep{chen2020}.  Note that the range of $\theta_k$'s in the Gaussian correlation function \eqref{eq:gscorr} should be non-negative, thus we re-write its form as 
\begin{equation}
	r(\bm x_j, \bm x_j)=\exp\left\{-\sum_{k=1}^d\phi_k^2|x_{ik}-x_{jk}|^2\right\},
\end{equation} 
where $\theta_k=\phi_k^2$. The range of $\phi_k$'s  is transformed into the all real axis and $\theta_k$ is remained to be non-negative. The  similar treatment is also considered in \citet{chen2020}.

The first stage of our stochastic search GP (SSGP) model is the GP model \eqref{eq:model} with $n$ observations $\mathcal{P}$ and $\bm y$, which can be expressed as a draw from the multivariate normal:
\begin{equation}\label{eq:sample}
	Y|\mu,\sigma^2, \bm\phi\sim N_n(\bm 1_n\mu, \sigma^2R(\bm\phi))
\end{equation}
where $\bm\phi=(\phi_1,\cdots,\phi_d)^{\mathrm{T}}$. And then the next stages specify the  prior distributions for parameters $\mu, \sigma^2, \bm \phi$. We first choose the non-informative priors for $\mu$ and $\sigma^2$ to simplify the computation, i.e., 
\begin{align}\label{eq:mu_prior}
	\pi(\mu)&\propto1\\
	\pi(\sigma^2)&\propto\frac{1}{\sigma^2}
\end{align}

Then let the prior for each $\theta_k$ be the normal mixture distribution: 
\begin{equation}\label{eq:theta_prior}
	\phi_k|\gamma_k\sim (1-\gamma_k)N(0, \tau_k^2)+\gamma_k N(0, c_k^2\tau_k^2)
\end{equation}
by introducing the latent variable $\gamma_k=0 \;  \text{or}\; 1$, s.t.,  
\begin{equation} \label{eq:gamma_prior}
	P(\gamma_k=0)=1-P(\gamma_k=1)=p_k
\end{equation}
to represent whether the value of  corresponding $\theta_k$ being $0$ or $1$, equivalently  whether the corresponding variable being  selected  or not.
That is, we set $\phi_k|(\gamma_k=0)\sim N(0, \tau_k^2)$ with small $\tau_k(>0)$ so that $\phi_k$ could be ``safely" estimated by $0$, and $\phi_k|(\gamma_k=1)\sim N(0, c_k^2\tau_k^2)$ with large $c_k(\gg1)$ so that the non-zero estimator of $\phi_k$ could be included. 
For convenience, we re-denote the prior for $\phi_k$ to the form of matrix:
\begin{equation}\label{eq:theta_matrix}
	\bm \phi|\bm \gamma \sim N(\bm 0, D_{\gamma}D_{\gamma}),
\end{equation} 
where $\bm \gamma=(\gamma_1,\cdots,\gamma_d)^{\mathrm{T}}$, $D_{\gamma}=\text{diag}\{\tau_1c_1^{\gamma_1}, \cdots,  \tau_dc_d^{\gamma_d}\}$.

Thus the stochastic search GP model is obtained  by combining \eqref{eq:sample}-\eqref{eq:theta_matrix}. And with independent priors assumption, the full posterior distribution of $\mu, \sigma^2, \bm \phi, \bm \gamma$ can be expressed as
\begin{equation}\label{eq:full_posterior}
    \begin{split}
	f(\mu, \sigma^2, \bm\phi, \bm\gamma|\mathcal{P}, \bm y) &\propto f(\bm y|\mu, \sigma^2, \bm\phi, \mathcal{P})\cdot \pi(\mu)\cdot \pi(\sigma^2)\cdot \pi(\bm \phi|\bm \gamma)\cdot \pi(\bm\gamma)\\
	&\propto \left(\det(\sigma^2R(\bm\phi)) \right)^{-\frac{1}{2}}\exp\left\{-\frac{1}{2\sigma^2}(\bm y- \bm1_n\mu)^{\mathrm{T}}R^{-1}(\bm \phi) (\bm y- \bm1_n\mu)\right\} \cdot  \\
	 & \quad 1 \cdot \frac{1}{\sigma^2} \cdot \left(\det(D_{\gamma}D_{\gamma}) \right)^{-\frac{1}{2}}\exp\left\{-\frac{1}{2}\bm\phi^{\mathrm{T}}(D_{\gamma}D_{\gamma})^{-1}\bm\phi \right\} \cdot \\
	 & \quad \prod_{k=1}^d p_k^{\gamma_k}(1-p_k)^{1-\gamma_k}
	\end{split}
\end{equation}
 
 Note that  our stochastic search GP model for variable selection in computer experiments is different from that of  \citet{linkletter2006} who used the ``spike and slab" mixture priors for correlation parameters that put a probability mass on $\theta_k=0$.  The distinction between these two approaches is analogous to that between \citet{george1993} and \citet{mitchell1988} in linear regression setup. And the difference between  \citet{huang2020} (named as SSBK) and our approach is straightforward that SSBK  selects the active terms for the mean function of universal Kriging and our approach selects variables through the Gaussian correlation function. 
 
\subsection{Posterior inference}
In the previous subsection,  the stochastic search GP model with normal mixture prior for correlation parameters which contains the information for variable selection, i.e., the marginal posterior distribution $\pi(\bm\gamma|\mathcal{P},\bm y)\propto f(\bm y|\mathcal{P},\bm\gamma)\pi(\bm \gamma)$, is specified. That is  $\pi(\bm\gamma|\mathcal{P},\bm y)$ provides a reference that can be used to select the more ``promising " subsets of  active variables. 
However, the full posterior distribution defined in \eqref{eq:full_posterior} is computationally intractable.
Markov Chain Monte Carlo algorithm is the most efficient and effective method to deal with this problem \citep{liang2011,robert2013}. Here, we use the Metropolis-within-Gibbs sampler to obtain a sequence $\bm\gamma^{(1)}, \cdots, \bm\gamma^{(M)}$ by generating an auxiliary ``Gibbs sequence":
\begin{equation} \label{eq:post_sequence}
	\mu^{(0)}, (\sigma^2)^{(0)}, \bm\phi^{(0)}, \bm\gamma^{(0)},\mu^{(1)}, (\sigma^2)^{(1)}, \bm\phi^{(1)}, \bm\gamma^{(1)}, \cdots, \mu^{(M)}, (\sigma^2)^{(M)}, \bm\phi^{(M)}, \bm\gamma^{(M)}
\end{equation}
where $\mu^{(0)}, (\sigma^2)^{(0)}, \bm\phi^{(0)}$ are initialised to be the MLEs of model \eqref{eq:model}, and $\bm\gamma^{(0)}$ is initialised to be $\bm\gamma^{(0)}=(1,1, \cdots,1)^{\mathrm{T}}$, the subsequent values of $\mu,\sigma^2,\bm\phi,\bm\gamma$ are successively simulating from the following sampling scheme, $M$ is the number of total iterations.

The samples of $\mu $ can be generated by
\begin{equation}\label{eq:post_mu}
	\mu|\sigma^2, \bm\phi,\bm\gamma, \mathcal{P}, \bm y\sim N\left((\bm1_n^{\mathrm T}R^{-1}(\bm\phi) \bm1_n)^{-1}\bm1_n^{\mathrm T}R^{-1}(\bm\phi) \bm y, \; \sigma^2(\bm1_n^{\mathrm T}R^{-1}(\bm\phi) \bm1_n)^{-1} \right);
\end{equation}
The samples of $\sigma^2$ can be generated by
\begin{equation}\label{eq:post_singma}
	\sigma^2|\mu,  \bm\phi,\bm\gamma, \mathcal{P}, \bm y \sim IG\left(\frac{n}{2}, \; \frac{(\bm y- \bm1_n\mu)^{\mathrm{T}}R^{-1}(\bm \phi) (\bm y- \bm1_n\mu)}{2} \right);
\end{equation}
Then, the full conditional distribution of $\bm\theta$ can be written as 
\begin{equation}\label{eq:post_theta}
\begin{split}
	\pi(\bm \phi|\mu, \sigma^2,\bm\gamma, \mathcal{P}, \bm y) &\propto  f(\bm y|\mu, \sigma^2, \bm\phi, \mathcal{P})\cdot\pi(\bm\phi|\bm\gamma)\\
	 &\propto \left(\det R(\bm\phi) \right)^{-\frac{1}{2}}\exp\left\{-\frac{1}{2\sigma^2}(\bm y- \bm1_n\mu)^{\mathrm{T}}R^{-1}(\bm \phi) (\bm y- \bm1_n\mu)-\frac{1}{2}\bm\phi^{\mathrm{T}}(D_{\gamma}D_{\gamma})^{-1}\bm\phi\right\};
\end{split}
\end{equation}
Because of its inexplicit formula, the Metropolis-Hastings scheme is embedded in the Gibbs sampler to generate samples from this full conditional distribution. We denote the density kernel of $\pi(\bm \phi|\mu, \sigma^2,\bm\gamma, \mathcal{P}, \bm y)$ to be $g(\bm\phi)$, and at step $(m+1)$, we simulate a candidate $\tilde{\bm\phi}$ as a alternative of  $\bm\phi^{(m)}$ by the  proposal density 
$$ q(\tilde{\bm\phi})= N(\bm\phi^{(m)}, \Sigma_{\bm\phi}),$$
 then we accept this sample $\tilde{\bm\phi}$ with the acceptance rate 
 $$\Lambda(\tilde{\bm\phi},\; \bm\phi^{(m)} )=\min\left\{1, \frac{g(\tilde{\bm\phi})}{g(\bm\phi^{(m)})} \right\}.
 $$
 This updating mechanism of $\bm \phi$ can be seen as a one-step random walk MH scheme. 
 
 Finally, we turn to parameters $\bm\gamma$.  Instead of directly sampling  the vector $\bm\gamma$, we sequentially update $\bm  \gamma$ componentwise by sampling from 
 \begin{equation}\label{eq:post_gamma}
 	\pi(\gamma_k| \mu, \sigma^2, \bm \phi,\gamma_{-k},\mathcal{P}, \bm y) = \pi(\gamma_k|\bm\phi,\gamma_{-k}) \propto \pi(\bm\phi|\gamma_k, \gamma_{-k}) \pi(\gamma_k, \gamma_{-k})
 \end{equation}
 where $\gamma_{-k}=(\gamma_1, \cdots, \gamma_{k-1}, \gamma_{k+1}, \cdots, \gamma_d)^{\mathrm T}$ denotes the vector of all $\gamma_k$'s except $\gamma_k$  . 
 Each distribution \eqref{eq:post_gamma} is Bernoulli with  probability 
 $$ P(\gamma_k=1|\bm\phi,\gamma_{-k})=\frac{a}{a+b},
 $$
 where 
 \begin{equation*}
 \begin{split}
 	a &= \pi(\bm\phi|\gamma_k=1, \gamma_{-k}) \pi(\gamma_k=1, \gamma_{-k}),\\
    b &= \pi(\bm\phi|\gamma_k=0, \gamma_{-k}) \pi(\gamma_k=0, \gamma_{-k})
 \end{split}
 \end{equation*}
 
 By sampling from \eqref{eq:post_mu} to  \eqref{eq:post_gamma} iteratively, the Gibbs sequence, an ergodic Markov chain \eqref{eq:post_sequence} is obtained.  And the information contained in the sequence  $\bm\gamma^{(1)}, \cdots, \bm\gamma^{(M)}$ relevant to variable selection for the GP model  can be extracted. In the meanwhile, following the posterior inference procedure described above,  the posterior mean,  the mean of MCMC simulation values of parameters $\mu, \sigma,\phi_k$ can be used to construct the GP model, and obtain the prediction \eqref{eq:km} and uncertainty quantification \eqref{eq:ks} for untried points.
 
 \subsection{Implement details}
The first issue that we should address in our SSGP model is the setup about the priors, especially for the hyper-parameters $c_i$ and $\tau_i$ for the correlation prior distribution. Tuning the hyper-parameters is critical for the model performance in Bayesian computation. The rough guide provided  by \cite{george1993}is that the small values of $\phi_k$'s should be within $3\tau_k$ with very high probability. Thus, following the discussion of related literatures \citep{huang2020,chen2020}, we suggest to set $\tau_k=1/(3\Delta \bm x)$, where $\Delta\bm x=(\max_i x_{i1}-\min_i x_{i1}, \cdots, \max_i x_{id}-\min_i x_{id})^{\mathrm T}, i=1,\cdots,n$. When the experimental domain is transformed into unit hypercube, $\tau_k$ is suggested to be $0.3$.  Generally speaking, the choice of $c_k$ is a fairly large positive value, $20$ or $25$ could be valid for most of our simulations. The uniform or ``indifference" prior is taken for $\bm\gamma$, i.e., $p_1=\cdots=p_d=0.5$, which means that each variable has an equal chance of being selected.        

A challenging issue when  programming MCMC algorithms is to determine whether the chain has approximately reached its stationary distribution. This problem known as convergence diagnosis is deemed as an art in Bayesian computation due to its reliance on the operators' experience.  There are many diagnostic tools such as simple graphical diagnostics and testing statistics \citep{givens2012,robert2013}. In our paper, the simple graphical diagnostics are adopted, for instance the samples trace diagram and auto-correlation plot. 

 After the sequence \eqref{eq:post_sequence} has reached approximate stationarity,  the values of $\bm \gamma$ corresponding to the most promising experimental factors wii appear with the highest frequency. Hence a simply tabulation method can be employed to show the high-frequency variables. In the meanwhile, the posterior estimation of parameters $\mu,\sigma^2$ and $\bm\phi$ for our SSGP model could be collected. Usually, the mean of posterior samples is taken. 
  
 In addition, in order to measure the ability of prediction, the empirical root mean squared prediction error (RMSPE) criterion \eqref{eq:RMSPE} and the median of absolute residuals (MAR) \eqref{eq:mar} are adopted. 
 
 \begin{equation}\label{eq:RMSPE}
 	\text{RMSPE}=\sqrt{\frac{1}{n_t}\sum_{i=1}^{n_t}\left(y(\bm x_i^*)-\hat{y}(\bm x_i^*)\right)^2}
 \end{equation} 
 
\begin{equation}\label{eq:mar}
	\text{MAR}=\text{median}\{|y(\bm x_i^*)-\hat{y}(\bm x_i^*)|: i=1.\cdots,n_t \}
\end{equation}
where $\bm x_i^*, \cdots, \bm x_{n_t}^*$ are the test points.  
 
\subsection{A  toy example}

In this subsection, a toy example modified from \cite{williams2006} is employed to elaborated our SSGP model for variable selection and parameters estimation. We consider modeling the output of the toy function
\begin{equation}
	f(\bm x)=(x_1^3+1)\cos(\pi x_2)+0\cdot x_3
\end{equation}
where $x_i\in [0,1]$ for all $i=1,2,3$. This function takes in a 3-dimensional input $\bm x=(x_1, x_2, x_3)^\mathrm{T}$, but the third variable $x_3$ does not have an impact on the output.  

First, a space-filling design, Latin hypercube design \citep{santner2018} or uniform design \citep{fang2018} is used to obtain the experimental data. For this toy example,  a 30-run Latin hypercube design with maximin criterion is evaluated. Then our SSGP model is used to analyse the data following the steps below: 
\begin{itemize}
	\item \textbf{Step 1}: Get the initial values $\mu^{(0)}=0.1302, (\sigma^2)^{(0)}=1.0387$ and $\bm\phi^{(0)}=(1.4567$, $1.7276, 0.5095)^{\mathrm{T}}$ from maximising the likelihood function. The maximum number of MCMC iterations is set to $6000$ and the first $2000$ draws are discarded as burn-in samples. 
	\item \textbf{Step 2}: Sample iteratively from \eqref{eq:post_mu} to  \eqref{eq:post_gamma}. This procedure is straightforward by Metropolis-within-Gibbs sampler.  In addition, some eye-ball convergence diagnosis could be done. In this toy example, the trace plot for parameters $\mu$ and $\sigma^2$ is shown in \autoref{fig:trace}.  

\begin{figure}[h]
	\centering
	\includegraphics[width=10.5cm, height=4cm]{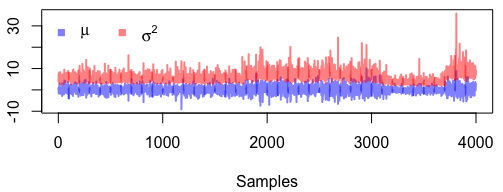}
	\caption{The trace plot for parameters in toy example}
	\label{fig:trace}
\end{figure}

	\item \textbf{Step 3}: Identify the important factors. By summarising the $4000$ valid simulations, we can tabulate the values of $\bm \gamma$.  The top three high-frequency values for $\gamma$ are listed in \autoref{tab:toy}.  It is shown that the active variables $x_1$ and $x_2$ are identified correctly with the highest frequency. 
\begin{table}[h]
	\centering
	\caption{Summary of the $\bm\gamma$ in toy example}
    \linespread{1.5}
    \label{tab:toy}
	\begin{tabular}{cccc}
	  \hline
    $\gamma_1$ & $\gamma_2$ & $\gamma_3$ & Frenquency   \\ \hline
    1 & 1 & 0 & $0.681$\\
    0 & 1 & 0 & $0.246$\\
    1 & 1 & 1 & $0.030$\\
     \hline
	\end{tabular}
\end{table}

\item \textbf{Step 4}: Predict the test data. For the purpose of prediction, 100 runs are generated randomly over the experimental domain $[0,1]^3$. And the posterior mean values of parameters $\mu, \sigma^2$ and $\bm\phi$ are taken, i.e., $\hat{\mu}=0.1340$, $\hat{\sigma^2}=5.9891$ and $\hat{\bm\phi}=(0.8795, 1.1072, -2.18\times 10^{-5})^{\mathrm T}$. Following the form of prediction (\ref{eq:km}), the predicted values for the 100 test points could be gotten. The RMSPE obtained by our SSGP model  is $0.00311$ and the value of MAR is $0.00057$, which are both better than the results ($0.02766$ and $0.01211$ respectively) by MLE method.  
\end{itemize}
This toy example demonstrates the detailed procedure of our SSGP model, The results show that this method is able to select the active variables with the highest frequency and obtain good prediction. More simulations are conducted in the next section.

\section{Numerical studies}
\subsection{Simulated examples}
In this subsection, simulations are conducted to compare our SSGP model with the method proposed by \cite{linkletter2006} (labeled as "RDVS").  The main idea of RDVS method is to use a reference distribution created by a known inert variable to identify  important factors through Bayesian analysis with a ``spike and slab" prior. The detailed procedure can be referred to \cite{linkletter2006} and Section 7.6 of \cite{santner2018}.   

\textit{Example} 1 (\textbf{test functions}). The simple linear  function and sinusoidal function modified from \cite{linkletter2006} are  
\begin{equation}
	\text{linaer:}\quad  f(\bm x )= 2x_1+ 2x_2+2x_3+2x_4;
\end{equation} 
\begin{equation}
	\text{sinusoidal:}\quad  f(\bm x )= \sin(x_1)+\sin(5x_2).
\end{equation} 
These two functions are augmented to be with $10$-dimensional input $\bm x=(x_1,x_2,\cdots,x_{10})^{\mathrm T}\in [0,1]^{10}$. The first four variables $x_1,x_2,x_3,x_4$ are deemed active and the remains are inactive for linear function and the first two $x_1,x_2$ are active for sinusoidal function.

A $54$-run Latin hypercube design is employed to obtain the output values for these two test functions.  The SSGP and RDVS programs are8 used to identify the important variables respectively. For the SSGP model, the maximum number of iterations is $6000$ including $2000$ burn-in samples. For RDVS,  $600$ MCMC simulations are adopted and the first 100 samples are discarded. $100$ repetitions are run to obtain the median distribution of the augmented inert variable and the upper $5, 10, 15\%$ quantiles of this reference distribution  play the role of baseline.    

Based on 100 repetitions, the simulation results are summarised in \autoref{tab:linear_result}. The column ``ACI" provides the average number of variables correctly identified (i.e. the average number of  the value of corresponding $\gamma_k$ being 1 for $k=1,2,3$ and $4$ in the  linear function) and the values in the parentheses, $\text{ACI}/4$, give the correct identification rate. The column ``AMI" gives the average number of variables misspecified (i.e. average number of the value of corresponding $\gamma_k$ being 1 for $k=5,\cdots,10$ in the linear function) and ``$\text{AMI/6}$" states the variable misspecification rate. The notations are analogous for sinusoidal function.  

\begin{table}[htbp]
	\centering
	\caption{ Comparisons based on the linear function}
    \linespread{1.5}
    \label{tab:linear_result}
	\begin{tabular}{cc|ccc|c}
	  \hline
   \multirow{2}*{Test function} & & \multicolumn{3}{c|}{RDVS} & \multirow{2}*{SSGP} \\ \cline{3-5}
 & & $5\%$ & $10\%$ & $15\%$  & \\ \hline
  \multirow{2}*{linear} &  ACI & 4 (1) & 4 (1) & 4 (1) & 4 (1) \\ 
  & AMI & 5.01 (0.84) & 3.15 (0.53) & 3.15 (0.19) & 0.36 (0.06) \\  \hline
  \multirow{2}*{sinusoidal} &  ACI & 2 (1) & 2 (1) & 2 (1) & 1.64 (0.82) \\ 
  & AMI & 6.66 (0.83) & 3.78 (0.47) & 1.73 (0.22) & 0 (0) \\  \hline
	\end{tabular}
\end{table}

The comparison results show that both SSGP and RDVS can identify the active factors correctly. However, SSGP selected less inert factors to be active than RDVS no matter which cut-off point is. On average, the misidentified number of the inactive factors is $0.36$ (far less than 1) for SSGP. In general, our proposed method performs well in terms of identifying the important variables.     

\vspace{1em}
\textit{Example} 2 (\textbf{borehole model}). The borehole model proposed by \cite{worley1987} is to describe the phenomenon about  the flow of the water through a borehole that is drilled from the ground surface through two aquifers.  Its simplicity and quick evaluation makes it a commonly used function for testing a wide variety of methods in computer experiments \citep{morris1993,xiong2013}. The  borehole function is 
\begin{equation}\label{eq:borehole}
	y_B(\bm x) = \frac{2\pi T_u(H_u-H_l)}{\ln (r/r_w) \left[1+\frac{2LT_u}{\ln(r/r_w)r_w^2K_w}+\frac{T_u}{T_w}  \right] }
\end{equation}
where the $\bm x$ inputs are listed in \autoref{tab:bore_input}.

\begin{table}[h]
	\centering
	\caption{Inputs and ranges of the borehole model (\ref{eq:borehole})}
    \linespread{1.5}
    \label{tab:bore_input}
	\begin{tabular}{cccc}
	  \hline
    Input  & Description & Units & Range  \\ \hline
    $r_w $ & Radius of the borehole  & m & $[0.05, 0.15]$\\
    $r$ & Radius of influence & m & $[100, 50000]$\\
    $ T_u$ & Transmissivity of the upper aquifer & $\text{m}^2/\text{year}$ & $[63070, 115600]$\\
    $H_u$ & Potentiometric head of the upper aquifer & m & $[990, 1100]$\\
    $T_l$ & Transmissivity of the lower aquifer & $\text{m}^2/\text{year}$ & $[63.1, 116]$\\
    $H_l$ & Potentiometric head of the lower aquifer & m &$[700, 820]$\\
    $L$ & Length of the borehole & m & $[1120, 1680]$\\
    $K_w$ & Hydraulic conductivity of borehole & $\text{m}/\text{year}$ & $[1500, 15000]$\\
     \hline
	\end{tabular}
\end{table}

As many researches (e.g., \cite{saltelli1995, joseph2008, moon2012, li2021}) mentioned that the borehole model itself is sparse.  The first and the eighth factors $r_w$ and $K_w$ are usually deemed important.  In this example, our SSGP model and the RDVS method are used to select the active variables and to predict the untried test experimental points. The MLE method programmed by R package ``DiceKriging" plays the role of benchmark for prediction.  

For SSGP model, the hyper-parameters are set as follows: $c_1=\cdots=c_8=15$ and $\tau_1=\cdots=\tau_8=0.3$. The maximum number of MCMC sampling is set to $5000$ and the first $1000$ draws are discarded. For RDVS,  an  known inactive variable is augmented randomly to the design matrix and $4000$ MCMC samples are simulated (including $2000$ burn-in samples).  This augmented procedure is repeated 100 times to obtain the reference distribution of the known inert variable, and the upper $15\%$ quantile of this reference distribution is taken as the baseline for RDVS.

Four common used space-filling designs with 50 runs are adopted to conduct the experiments and to examine the performance of our method. They are Latin hypercube design with maximin criterion, Latin hypercube design with maximum projection criterion,  uniform design with modified $L_2$-discrepancy and rotated sphere packing design. 

\begin{table}[hbtp]
	\centering
	\caption{Results for the borehole model}
    \linespread{1.5}
    \label{tab:borehole}
    \begin{threeparttable}
	\begin{tabular}{c|ccc|c|ccc}
	  \hline
\multicolumn{4}{c|}{LHD (maxmin)} & \multicolumn{4}{c}{LHD (maxpro)} \\ \hline
Method & A.V. & RMSPE & MAR &  Method & A.V. & RMSPE & MAR \\ \hline
MLE & --- & 5.2684 & 2.0684 & MLE &  --- & 4.9762 & 2.5001 \\ 
RDVS  & $r_w,H_u, L, K_w$  & 20.3088 & 7.7595 & RDVS & $r_w, H_l, K_w$ & 16.9663 & 7.4075 \\ 
SSGP & $r_w, K_w$ & 1.9595 & 0.6532 & SSGP & $r_w$ & 2.1098 & 0.9903 \\
     \hline
 \multicolumn{4}{c|}{UD ($MD_2$)} & \multicolumn{4}{c}{RSPD} \\ \hline
 Method & A.V. & RMSPE & MAR &  Method & A.V. & RMSPE & MAR \\ \hline
MLE & --- & 5.0076 & 2.4986 & MLE &  --- & 5.3465 & 3.0076 \\ 
RDVS  & $r_w,H_u, H_l, K_w$  & 20.7681 & 8.0807 & RDVS & $r_w, H_l, K_w$ & 16.6786 & 7.8572 \\ 
SSGP & $r_w, K_w$ & 1.9908 & 0.9564 & SSGP & $r_w$ & 1.7792 & 0.8342 \\
\hline
	\end{tabular}
	\begin{tablenotes}
	\linespread{1}
        \footnotesize
        \item[1] ``LHD (maximin)" means the Latin hypercube design with maximin criterion; ``LHD (maxpro)" means Latin hypercube design with maximum projection criterion; ``UD ($MD_2$)" means the uniform design with modified $L_2$-discrepancy; ``RSPD" means the rotated sphere packing design.
        \item[2] ``A.V." means the active variables identified by the corresponding method. ``---" corresponding to MLE means this method is not for variable selection. 
    \end{tablenotes}
\end{threeparttable}
\end{table}

The simulation results are listed in \autoref{tab:borehole}, including the important variables identified by RDVS and SSGP and the RMSPE and MAR values for test points.  No matter what designs are selected, RDVS includes other variables as the active inputs except for $r_w, K_w$.   However, our SSGP can identify this two factors, $r_w$ and $K_w$ as the important variables under the case of Latin hypercube design with maximin criterion and uniform design with modified $L_2$-discrepancy, which is in line with other screening methods mentioned above.  In addition, $500$ test points are generated randomly to measure the prediction performance for SSGP, RDVS and MLE method. The results indicate that SSGP performs much better than RDVS and MLE  under all the four space-filling designs.  

\subsection{A practical example}

Piston slop is an unwanted engine noise caused by piston secondary motion. Power cylinder system is modeled using the multibody dynamics code ADAMS/Flex, which also includes a finite-element model \citep{hoffman2003}.  The computer experiment is employed to explore the relationship between piston noise and its covariates,  that are set clearance between the piston and the cylinder liner ($x_1$), location of peak pressure ($x_2$), skirt length ($x_3$), skirt profile ($x_4$), skirt ovality ($x_5$), and pin offset ($x_6$). More detailed description about this practical problem can be refereed to \cite{li2005,fang2006} and the references therein.

A uniform design with $12$ runs for this piston slab experiment is adopted and the collected data provided by \cite{li2005} is displayed  in \autoref{tab:piston_data}. Then our SSGP model is used to analyse the piston  slap noise data. Following our advice about the hyper-parameters for the prior,  we set $c_1=\cdots=c_6=0.3$, $\tau_1=\cdots=\tau_6=25$ and $\Sigma_{\bm \phi}=\text{diag}\{0.03 ,\cdots, 0.03\}$. The process of  MCMC sampling runs 7000 iterations including 1500 burn-in samples. 
\begin{table}[htbp]
	\centering
	\caption{Piston slap noise data}
    \linespread{1.5}
    \label{tab:piston_data}
	\begin{tabular}{cccccccc}
	  \hline
 Run $\#$ & $x_1$ & $x_2$ & $x_3$ & $x_4$ & $x_5$ & $x_6$ & Noise(dB)  \\ \hline
1 & 71 & 16.8 & 21.0 & 2 & 1 & 0.98 & 56.75 \\
2 & 15 & 15.6 & 21.8 & 1 & 2 & 1.30 & 57.65 \\
3 & 29 & 14.4 & 25.0 & 2 & 1 & 1.14 & 53.97 \\
4 & 85 & 14.4 & 21.8 & 2 & 3 & 0.66 & 58.77 \\
5 & 29 & 12.0 & 21.0 & 3 & 2 & 0.82 & 56.34 \\
6 & 57 & 12.0 & 23.4 & 1 & 3 & 0.98 & 56.85 \\
7 & 85 & 13.2 & 24.2 & 3 & 2 & 1.30 & 56.68 \\
8 & 71 & 18.0 & 25.0 & 1 & 2 & 0.82 & 58.45 \\
9 & 43 & 18.0 & 22.6 & 3 & 3 & 1.14 & 55.50 \\
10 & 15 & 16.8 & 24.2 & 2 & 3 & 0.50 & 52.77 \\
11 & 43 & 13.2 & 22.6 & 1 & 1 & 0.50 & 57.36 \\
12 & 57 & 15.6 & 23.4 & 3 & 1 & 0.66 & 59.64 \\
     \hline
	\end{tabular}
\end{table}

After MCMC simulation finished, the values of $\bm\gamma$ is summarised.  The variables $x_1$, $x_5$ and $x_6$ are identified with the highest frequency $48.49\%$. In addition, in order to assess the prediction performance of our method,  an additional 100 testing points that are available from the website \url{http://www.personal.psu.edu/ril4/DMCE/MatlabCode/} is adopted. The results are listed in \autoref{tab:piston_result}, including those through penalised method reported by \cite{fang2006}.  The comparison indicates that SSGP can perform better than the penalised likelihood method (including $l_1$, $l_2$ and SCAD penalty).

\begin{table}[htbp]
	\centering
	\caption{ Prediction results in piston slap}
    \label{tab:piston_result}
	\begin{tabular}{c|c|cc}
	  \hline
 \multicolumn{2}{c|}{Method} &  RMSPE & MAR \\ \hline
 \multicolumn{2}{c|}{MLE} & 1.7307 & 1.3375 \\ \hline
\multirow{3}*{\tabincell{c}{Penalised \\ likelihood}} & $l_1$ & 2.0753  & 1.4638 \\
& $l_2$ & 1.7632  & 1.3114 \\
& SCAD & 1.4813 & 1.0588 \\ \hline
\multicolumn{2}{c|}{SSGP} & 1.3573 & 1.0554 \\
     \hline
	\end{tabular}
\end{table}

\section{Conclusion and discussion}
This work is focused on the  variable selection issue  for Gaussian process model which is a critical problem in computer experiments.  A new Bayesian Gaussian process model is constructed which stem from the idea of  Stochastic Search Variable Selection (SSVS) in \cite{george1993}. The main thrust of our proposed model is to adopt the normal mixture priors for the correlation parameters and latent indicator variables are embedded into these priors for the sake of identifying active factors.  The goals of variable selection and parameter estimation have been achieved through the efficient MCMC algorithm---Metropolis-within-Gibbs sampler. The empirical results in simulated examples including a test function and the borehole model show that our method performs much better than the RDVS procedure in the terms of accuracy of selecting important variables and prediction for untried points.  A real application to piston slap noise data was also programmed and our model can obtain better prediction than the penalised likelihood method. 

 A few issues should be paid attention to when operating our method. First, the choice of prior for the correlation parameters. Because the parameter $\theta_k$'s are non-negative in Gaussian correlation function \eqref{eq:gscorr}, the common prior for them in Bayesian Gaussian process is the gamma distribution. The natural try is to use a mixture of gamma distribution for the sake of variable selection, However this idea can not work stably in various of problems. Thus the strategy of reparameterization is considered in our paper. The range of the correlation parameter is transformed into the all real numbers and the normal mixture prior for variable selection can be adopted. Another issue is the high computational costs of MCMC algorithm. The technique of variational inference could be an alternative in our future research.        

\section*{Acknowledgements}
This study is partially supported by the National Natural Science Foundation of China (No.11571133 and 11871237).

\bibliographystyle{plainnat}
\bibliography{ref}   

\end{document}